\documentclass{svproc}
\usepackage{url}
\usepackage{epsfig}

\begin{document}
\mainmatter              
\title{Patient-specific immobilisation for radiotherapy treatment of extensive lower-limb carcinoma}
\titlerunning{Patient-specific immobilisation of lower limb}  

\author{Tanya Kairn\inst{1,2,3} \and Tania Poroa\inst{1,4} \and Jenna Luscombe\inst{1,4} \and Shaun Chan\inst{1} \and Michelle Grogan\inst{1,2} \and Scott B Crowe\inst{1,2,3,4}}

\authorrunning{Tanya Kairn et al.}
\tocauthor{Tanya Kairn, Tania Poroa, Jenna Luscombe, Shaun Chan, Michelle Grogan, and Scott B Crowe}

\institute{Royal Brisbane and Women's Hospital, Brisbane, Qld, Australia\\
\and
University of Queensland, Brisbane, Qld, Australia
\and
Queensland University of Technology, Brisbane, Qld, Australia
\and
Herston Biofabrication Institute, Metro North Hospital and Health Service, Brisbane, Qld, Australia\\
\email{t.kairn@gmail.com}}

\maketitle              

\begin{abstract}
When non-melanoma skin cancers extend over large areas of skin, effective radiotherapy treatments can be delivered using volumetric modulated arc therapy (VMAT) beams with narrow segments that rotate around the affected surfaces. For lower-limb carcinoma treatments, careful immobilisation is needed to achieve the degree of reproducible and stable positioning required to ensure the narrow field segments treat the targeted tissue and avoid underlying anatomy. To meet this need, a 3D printed patient-specific foot support was created in the form of a solid box containing a deep “footprint”, shaped to match the outline of the patient’s relaxed foot when lying supine, supported by a vacuum bag. For our first patient treated with a 3D printed foot support, image guidance data and clinical notes were evaluated against corresponding information from all recent lower-leg VMAT treatments. The patient feedback was recorded as “good”, with no complaints of discomfort or poor fit, and the proportion of treatment fractions requiring shifts greater than 5 mm after setup imaging (20\%) compared favourably to the proportions for patients without a patient-specific foot support (23\%-84\%). The patient-specific foot support was particularly useful for minimising longitudinal shifts and leg rotations. Evidently, 3D-printed patient-specific immobilisation devices have the potential to enhance positioning stability, and therefore potentially improve accuracy and effectiveness of VMAT treatments, for patients with extensive lower-limb carcinomas.

\keywords{Radiation therapy, additive manufacture, fused deposition modelling, immobilisation}
\end{abstract}

\section{Introduction}

Volumetric modulated arc therapy (VMAT) techniques can be used to effectively control otherwise challenging non-melanoma skin cancers that extend over large areas of skin \cite{chua2019,wills2021}. VMAT treatments that deliver tumoricidal doses to large, convex superficial regions while sparing underlying tissues often involve tangential delivery of long, narrow beam segments \cite{spelleken2022,nakayama2021} that require particular attention to patient setup and immobilisation.

While the use of vacuum bags for limb immobilisation is well established for static conformal radiotherapy treatments \cite{swinscoe2018}, the increased prevision required when using VMAT to specifically target the skin and superficial tissues has led to a growth of work in the area of adapting and re-purposing thermoplastic shells \cite{swinscoe2018,simoes2020}. For example, Zheng et al have reported on their modifications of commercial U-shaped thermoplastic masks, designed for cranial immobilisation, to allow application to knees, feet and lower legs \cite{zheng2012}.

As an alternative approach to the challenge of lower-limb immobilisation, this study investigated the fabrication and use of a 3D printed patient-specific foot support in combination with a standard vacuum bag. For our first patient treated with a 3D printed foot support, image guidance data and clinical notes were evaluated against corresponding information from all recent lower-leg VMAT treatments, to investigate the potential effects of this intervention on treatment setup accuracy.

\section{Method}

\vspace{-0.5cm}
\begin{figure}[htbp]
\begin{center}
\includegraphics[width=0.9\linewidth]{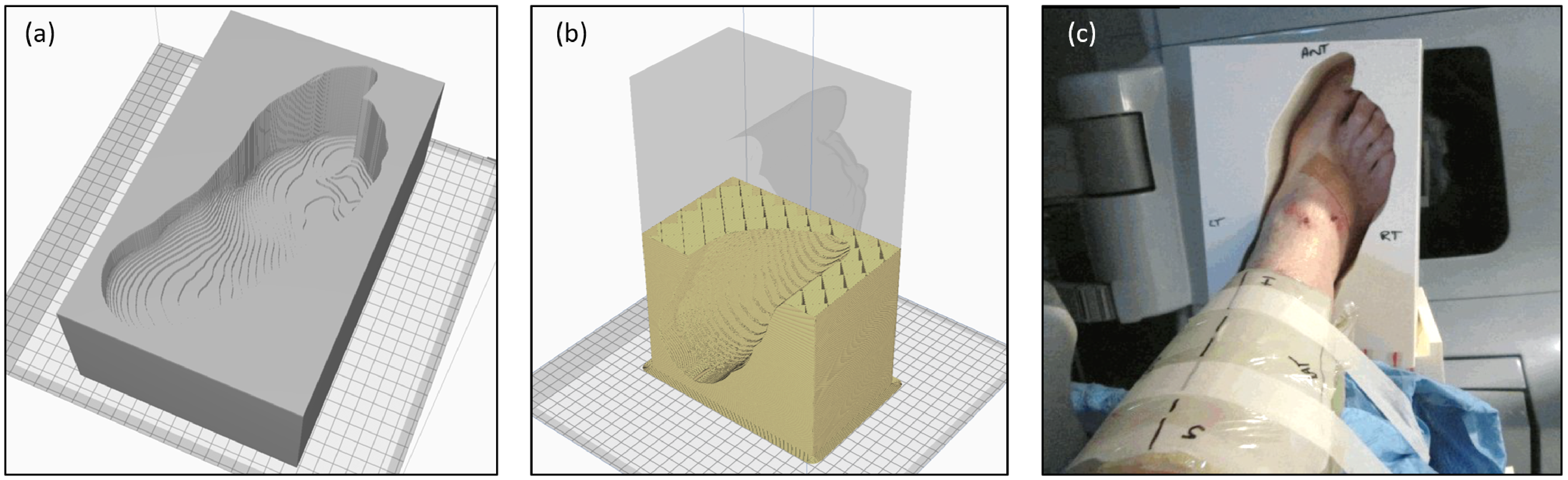}
\caption{Patient-specific foot support shown as (a) a 3D render of the ``footprint'' design, (b) coss-section showing internal structure of the print and (c) a photograph showing the use of the finished support.}
\label{fig:1}
\end{center}
\end{figure}
\vspace{-0.8cm}

This study involved the aggregation and retrospective analysis of image guided radiotherapy (IGRT) data from five cases with extensive and multiple carcinomas of the skin and superficial tissues of the lower leg; two cases of basal cell carcinoma, two cases of squamous cell carcinoma and one case of Merkel cell carcinoma. All patients were treated in the feet-first supine position, using personalised vacuum bags for positioning and immobilisation of the lower legs. Of these five treatments, one (a basal cell carcinoma case) was planned and delivered with the vacuum bag immobilisation augmented by a 3D printed patient-specific foot support.

The patient-specific foot support was designed with reference to a CT image using MIM Maestro structure contouring tools (MIM Software Inc, Cleveland, USA). A solid box containing a deep ``footprint'' (see figure \ref{fig:1}(a)) was carefully shaped to match the outline of the patient’s relaxed foot when lying supine in their treatment position. 

The support was printed using polylactic acid (PLA) filament on an Ultimaker 2+ 3D printer (Ultimaker BV, Geldermalsen, Netherlands). Although the foot support was located beyond the path of the radiation beam, a low in-fill density (5\%) was used to minimise any possible scattering effects, while a regular grid pattern was used to achieve a mechanically robust print (see figure \ref{fig:1}(b)). Printing of the support (shown in figure \ref{fig:1}(c)) took 12 hours to complete. 

The treatments for all five cases examined in this study were planned with a volumetric modulated arc therapy (VMAT) technique using the Varian Eclipse treatment planning system, with nominal a 6 MV photon beam delivered using a Varian iX linac (Varian Medical Systems, Palo Alto, USA). Dose build-up at the skin was achieved using a gel bolus. 

Image guidance was performed using orthogonal pairs of two-dimensional kV x-ray images, via the Varian On Board Imaging (OBI) system. The OBI system and the iX linac couch had a precision of 1 mm and automated couch corrections were limited to the longitudinal, lateral and vertical axes plus yaw rotation. Uncorrected pitch and roll rotations were logged manually alongside the other shifts identified by the OBI system and these results were routinely uploaded into the oncology information system from where they were accessed (along with clinical notes and other data) for analysis in this study.

This retrospective review of clinical data was approved by the Royal Brisbane and Women's Hospital Human Research Ethics Committee (RBWH HREC, EC00172) and this study was completed in accordance with the ethical standards of the RBWH HREC and the 1964 Helsinki declaration and its later amendments.

\section{Results}

\begin{figure}[htbp]
\begin{center}
\includegraphics[width=0.8\linewidth]{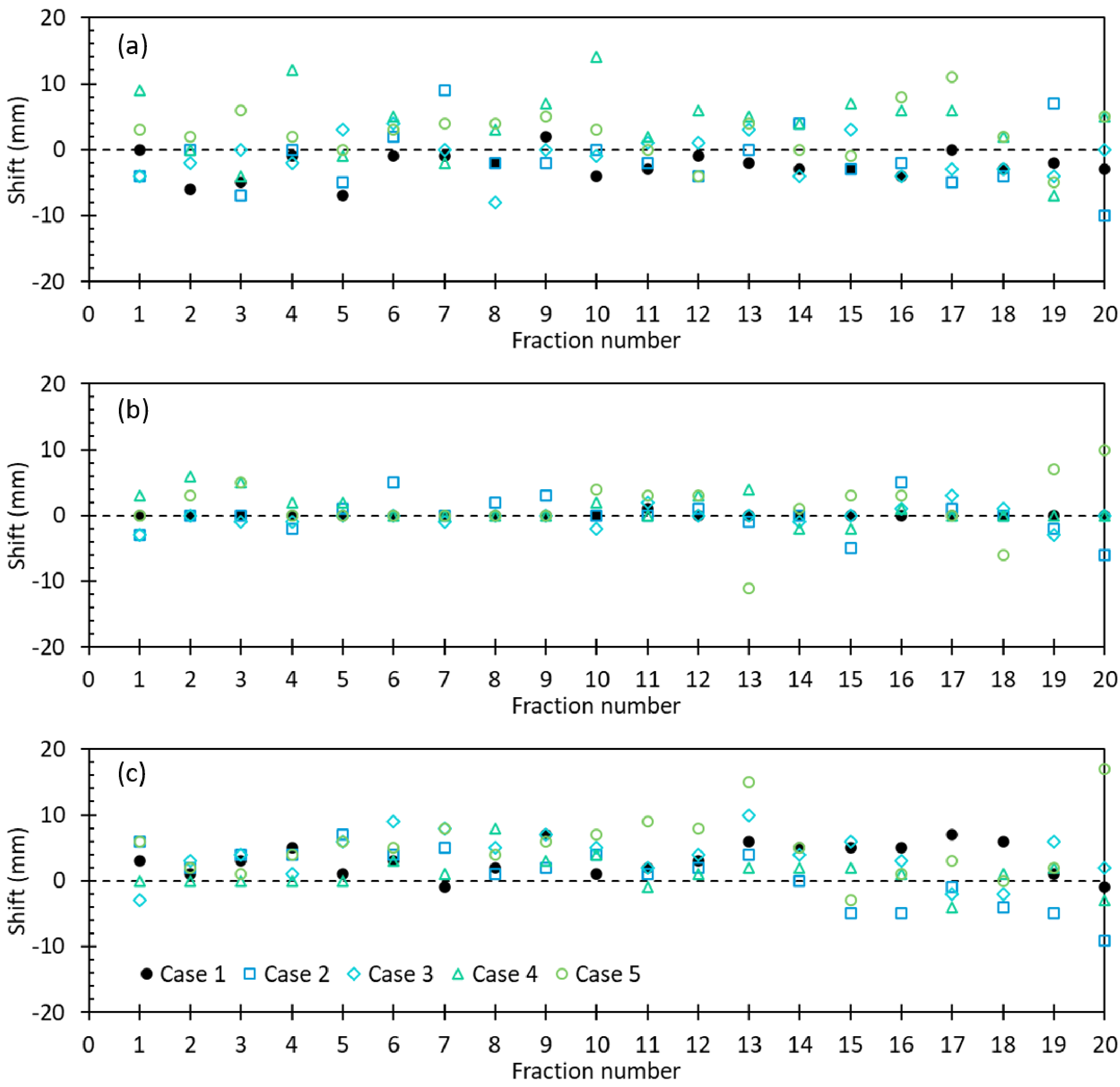}
\caption{Orthogonal shift corrections identified by IGRT system in the (a) lateral, (b) longitudinal and (c) vertical directions. Cases are labelled 1 to 5 in arbitrary order, with case 1 being the treatment that used the patient-specific foot support}
\label{fig:2}
\end{center}
\end{figure}

The performance of the support in assisting with lower-limb setup and immobilisation is demonstrated by the reduction in orthogonal shift corrections, seen in figures \ref{fig:2}(a)-(c), in comparison to the four cases where vacuum bags were used for lower-limb immobilisation without additional 3D printed supports. In particular, results in figure \ref{fig:2}(b) show how the use of the patient-specific foot support in combination with the indexing of the treatment couch has effectively eliminated the need for longitudinal position corrections, in this case.

Data in figures \ref{fig:2}(a)-(c) also indicate that for the cases where no patient-specific foot supports were used, the IGRT system identified correction shifts greater than 5 mm in 23-84\% of fractions and shifts greater than 10 mm in 3-20\% of fractions. By comparison, for the case where the patient-specific foot support was used, shifts greater than 5 mm were needed in only 20\% of fractions and shifts greater than 10 mm were not required for any fractions. 

Similarly, substantial uncorrected pitch and roll rotations (generally 1-2$^\circ$) were noted affecting 8-20\% of the treatment fractions delivered in the cases without patient-specific foot supports and none of the treatment fractions delivered to the patient with the 3D printed support. The patient's response to the 3D printed foot support was recorded as ``good'', with no complaints regarding comfort or fit. 

\vspace{-0.5cm}
\vspace{-0.8cm}

\section{Discussion}

The results of analysing IGRT data for these five cases suggest that the use of patient-specific supports to augment the conventional vacuum immobilisation of lower-limb radiotherapy patients has the potential to improve treatment setup accuracy and thereby enhance treatment delivery reliability. In particular, no longitudinal shift corrections greater than 1 mm or uncorrected rotations were recorded in the case where the patient was treated with a carefully fitted 3D printed foot support. 

These promising results were achieved despite this study being limited to only a small number of cases, due to the infrequency with which our department sees extensive lower-limb carcinomas needing radiotherapy treatments. It is therefore inadvisable to extrapolate these observed effects on treatment setup accuracy to the entire lower-limb carcinoma patient population. Positive effects may be enhanced or degraded by numerous factors including patient height, weight, comorbidities and compliance, as well as the location or severity of the treated lesions. Further studies are needed to examine these effects and identify the patient cohorts most likely to benefit from this intervention.

\section{Conclusion}

This small study provides a useful proof-of-concept and example method for creating patient-specific 3D printed foot supports to potentially enhance positioning reliability, and therefore achieve improved accuracy and effectiveness of radiotherapy treatments, for patients with extensive lower-limb carcinomas. 

\section*{Acknowledgements}
Contributions to this work from Tania Poroa, Jenna Luscombe and Scott B. Crowe were supported by a Metro North Hospital and Health Service funded Herston Biofabrication Institute Programme Grant.

\end{document}